\documentclass[twocolumn,10pt]{IEEEtran}
\usepackage{comment}
\includecomment{toexclude}
\usepackage{amsmath,amsfonts,amssymb}
\usepackage{cite}
\usepackage{verbatim}
\usepackage{bm}
\usepackage{algorithm}
\usepackage{graphicx}
\usepackage{xcolor}
\usepackage{subcaption}
\usepackage[colorlinks=true,citecolor=blue,linkcolor=blue]{hyperref}

\graphicspath{{./images/}}

\newtheorem{thm}{Theorem}
\newtheorem{rmk}{Remark}

\newtheorem{def1}{Definition}
\newtheorem{lem}{Lemma}
{}

\newtheorem{cor1}{Corollary}

\title{Stability Analysis for Switched Systems with Sequence-based Average Dwell Time}

\author{Dianhao~Zheng, Hongbin~Zhang,~\IEEEmembership{Senior~Member,~IEEE},
	J. Andrew Zhang,~\IEEEmembership{Senior~Member,~IEEE},
	and~Steven~W.~Su,~\IEEEmembership{Senior~Member,~IEEE}
	\thanks{Manuscript received xxxx; revised xxx; accepted xxxx. This
		work was supported in part by National Natural Science Foundation of China (Grant no. 61374117). Recommended by xxxx. Corresponding author: xxxxx}
	\thanks{Dianhao Zheng is with the School of Information and Communication Engineering, University of Electronic Science and Technology of China, Chengdu 611731, China, is also with the Faculty of Engineering and Information Technology, University of Technology Sydney, Ultimo, NSW 2007, Australia (e-mail: dianhao18@126.com).}
	\thanks{Hongbin Zhang is with the School of Information and Communication, University of Electronic Science and Technology of China, Chengdu 611731, China (e-mail:zhanghb@uestc.edu.cn).}
	\thanks{J. Andrew Zhang is with the Global Big Data Technologies Centre, University of Technology Sydney, Ultimo, NSW 2007, Australia (e-mail:Andrew.Zhang@uts.edu.au).}
	\thanks{Steven W. Su is with the Faculty of Engineering and IT, University  of Technology  Sydney, Ultimo, NSW 2007, Australia (e-mail:Steven.Su@uts.edu.au)}
}

\begin{document}

\maketitle

\begin{abstract}
This note investigates the stability of both linear and nonlinear switched systems with average dwell time. Two new analysis methods are proposed. Different from existing approaches, the proposed methods take into account the sequence in which the subsystems are switched. Depending on the predecessor or successor subsystems to be considered, sequence-based average preceding dwell time (SBAPDT) and sequence-based average subsequence dwell time (SBASDT) approaches are proposed and discussed for both continuous and discrete time systems. These proposed methods, when considering the switch sequence, have the potential to further reduce the conservativeness of the existing approaches. A comparative numerical example is also given to demonstrate the advantages of the proposed approaches.
\end{abstract}


\section{Introduction}\label{sec-intro}

Switched systems consist of a finite number of subsystems and a logical rule that orchestrates switching between these subsystems \cite{lin2009stability}, and are often used to model many physical or man-made systems\cite{zhang2010asynchronously,zhao2012stability}. In recent years, considerable research efforts have been devoted to the study of the switched system \cite{fei2018quasi,wu2017mode,wu2018concurrent,fei2017asynchronous}, for its wide range of applications, such as cooperation of multi-agent systems\cite{zheng2016consensus,zheng2018consensus}, modeling of electronic switching systems.

The stability of dynamical systems \cite{lasalle1976stability,praly1986global} is a broad area in system and control theory. In recent years, the stability analysis of switched systems becomes one of the active research topics.
In 1996, paper \cite{ morse1996supervisory} presented a definition on dwell time, being set as a constant, to solve the problem of supervisory control of families of linear controllers.
Three years later, a more general and flexible definition for the switching signal, the signal with average dwell time (ADT) switching, was proposed in \cite{hespanha1999stability}. This scheme requires that the $average$ interval between consecutive discontinuities is not less than a constant.
On the basis of this innovative idea, a number of useful techniques have been developed  \cite{zhang2010asynchronously,lian2010sliding,zhang2009stability,zong2015finite}.
In 2011, paper \cite{zhao2012stability} proposed a new concept for analysis of the stability of switched systems, mode-dependent average dwell time (MDADT), which could reduce the restrictions on the average dwell time (ADT).
The key idea is the specification, in which $each$ $mode$ in the underlying system has its own ADT. On the basis of the concept of MDADT, several new research results have been reported \cite{fei2018quasi,wu2017mode,wu2018concurrent,fei2017asynchronous}.

However, improving stability analysis for switched systems is far from completion yet, even in the linear context. To be more specific, for existing methods, if the settings of the switching scenarios can be further refined, the conservativeness of the switching law, e.g., the average dwell time, could be further reduced. Under this consideration, to relax the restrictions of the switching law, motivated by \cite{zhao2012stability}, we not only consider the ADT for each $individual$ $mode$, but also the $predecessor$ and $successor$ of each switching mode. That is, rather than just considering the ADT for each mode, we further investigate the ADT for each mode with their $switching$ $sequence$ taking into account.

In this study, we propose two new methods for analyzing the stability of linear and nonlinear switched systems and design the corresponding feedback controllers in both continuous and discrete time situations.
 The major contributions of this note can be summarized as follows:
1) Two new concepts have been originally proposed and defined, namely the sequence-based average subsequence dwell time and the sequence-based average preceding dwell time.
2) We develop two new methods, which distinguish subsequence ADT from preceding ADT and propose parallel stability conditions for switched linear and nonlinear systems.
3) On the basis of the proposed new methods, we derive two less conservative stability conditions, which could further reduce the mode-dependent average dwell time compared to the existing approaches. Furthermore, we show that many existing methods can be generalized as special and simplified versions of the proposed methods.
Finally, an illustrative example is provided, and the effectiveness of the proposed approaches are demonstrated.

The rest of this note is organized as follows. In Section II, some basic concepts and lemmas are introduced. In Section III two sequence-based average dwell time methods are proposed and applied to linear systems. Section IV provides an example to verify the obtained results in this note.

\textit{Notation:} The notations used in this note are very standard. The symbol "$\times$" represents the multiplication operation or Cartesian product of sets. The symbol $\rightarrow 0$ means approaching zero. Throughout this note, for a given matrix $P$, $P>(<)~0$ implies that the matrix $P$ is symmetric and positive (or negative) definite. A superscript ``T" stands for the matrix transpose, and a function $\kappa$ is said to be of class $\mathcal{K}_{\infty }$ function if the function $\kappa: [0,\infty)\rightarrow[0,\infty)$, $\kappa(0) = 0$, is strictly increasing, continuous, unbounded. The time $t_1,{{t}_{2}},{{t}_{3}},\cdots,t_l,t_{l+1},\cdots $ or $k_1,k_2,..,k_l,k_{l+1}$,..., are the switching times of subsystems. The flags $t_l^-$ and $t_l$ stand for the time just before or after subsystems switching at time $t_l$. The function $\sigma (t):[0, + \infty ) \to {\mathcal S} = \{ 1,2, \cdots ,s\} $  is the switching signal and $s$ is the total amount of subsystems. The mark $p|q$ stands for the case that the $p^{th}$ subsystem is activated immediately after the $q^{th}$ subsystem.

 \section{PRELIMINARIES}

 In this section, we first recall some concepts about switched systems and then we review some recent related results.

 It is important for switched systems to reach a stable status. In order to describe the stability of switched systems, we first give the definition of global uniformly exponential stable (GUES).

 \begin{def1}\cite{zhao2012stability}\label{p8_def1}
 	For a given system and a switching signal $\sigma(t)$, an equilibrium $ x^*=0$ is globally uniformly exponentially stable (GUES) if there exist constants $\eta >0$, $\gamma>0$ (or $0<\varsigma<1$) for any initial conditions $x(t_0)$ (or $x(k_0)$), such that the solution of the system satisfies $\parallel\! x(t)\!\parallel \leq \eta \parallel\! x(t_0)\! \parallel e^{-\gamma(t-t_0)},\forall t\geq t_0 $ for the continuous-time situation (or $\parallel\! x(k)\!\parallel \leq \eta \parallel\! x(k_0)\! \parallel \varsigma^{(k-k_0)}, \forall k\geq k_0$ for the discrete-time case).
 \end{def1}

 In order to analyse the stability of switched systems, \cite{zhao2012stability} and \cite{hespanha1999stability} proposed important concepts as follows.

 \begin{def1}\cite{zhao2012stability}\label{p8_def2}
 	For any switching times $t_2 \ge t_1 \ge 0 $ with a switching signal $\sigma(t)$, let $N_{\sigma p}(t_2,t_1)$ stand for the number of the $p^{th}$ subsystem that is activated over the time interval $[t_1,t_2]$, and $T_{p}(t_2,t_1)$ denote the total running time of the $p^{th}$ subsystem over the time interval $[t_1,t_2]$, $p \in \mathcal{S}$. It is said that $\sigma(t)$ has a mode-dependent average dwell time $\tau_{ap}$ if there
 	exist two subsequence numbers $N_{0 p}$ ($N_{0 p}$ is called the mode-dependent chatter bound here) and $\tau_{ap}$ such that
 	\[
 	N_{\sigma p} (t_2,t_1)\le N_{0p} +\frac{T_{p}(t_2,t_1)}{\tau_{ap} }.
 	\]
 \end{def1}

 Consider a class of continuous-time or discrete-time switched linear systems:

\begin{equation}
 \label{p8_e1}
 \begin{array}{c}
  \dot x(t)=A_{\sigma (t)}x(t)+B_{\sigma (t)}u(t),
 \end{array}
 \end{equation}

 \begin{equation}
 \label{p8_e2}
 \begin{array}{c}
  x(k+1)=A_{\sigma (k)}x(k)+B_{\sigma (k)}u(k)
 \end{array}
 \end{equation}
where $x(t)$ ($x(k)$) $\in \mathcal R^n$ and $u(t)$ ($u(k)$) are the state vector and the control input.

For systems (\ref{p8_e1}) and (\ref{p8_e2}), some useful results can be found in recent literature.

 \begin{lem}\cite{zhao2012stability}(Continuous-time Situation)\label{p8_lem1}
 	For the system (\ref{p8_e1}) when $u(t)\equiv0$ and given constants $\lambda_p >0$, $\mu_p>1$, $p\in \rm{{\mathcal S}}$. Suppose that there exist matrices $P_p>0$, such that $\forall p\in \rm{{\mathcal S}}$,
 	$A_p^TP_p+P_pA_p\le - \lambda_p P_p$
 	and $\forall (p,q)\in {\mathcal S}\times {\mathcal S},p\ne q$,  it holds that $P_p \le \mu_p P_q.$
 	Then the system is globally uniformly asymptotically stable with the MDADT condition
 	\begin{equation}
 	\label{p8_e3}
 	\tau_{ap} \ge \tau_{ap}^\ast=ln\mu_p/\lambda_p.
 	\end{equation}
 	
 \end{lem}

 \begin{lem}\cite{zhao2012stability}(Discrete-time Situation)\label{p8_lem2}
 	For the system (\ref{p8_e2}), when $u(k)\equiv0$, and the given constants $1>\lambda_p >0$, $\mu_p\ge1$, $p\in \rm{{\mathcal S}}$, suppose that there exist matrices $P_p>0$ such that $\forall p\in \rm{{\mathcal S}}$,
 	$A_p^TP_pA_p-P_p\le - \lambda_p P_p$
 	and $\forall (p,q)\in {\mathcal S}\times {\mathcal S},p\ne q$,  it holds that
 	$	P_p \le \mu_p P_q.$
 	Then the system is globally uniformly asymptotically stable with the MDADT condition
 	\begin{equation}
 	\label{p8_e4}
 	\tau_{ap} \ge \tau_{ap}^\ast=-ln\mu_p/(1-\lambda_p).
 	\end{equation}
 \end{lem}

 In the situation of $u(t)\not\equiv 0$, the work \cite{zhao2012stability} considered a class of feedback $u(t)=K_{\sigma(t)}x(t)$ for the continuous-time case or $u(k)=K_{\sigma(k)}x(k)$ for the discrete-time case, where $K_p$, $p\in {\mathcal S}$, is the controller gain. The closed-loop system is given by
 \begin{equation}
 \label{p8_e5}
 \begin{array}{c}
 \dot{x}(t)=(A_{\sigma (t)}+B_{\sigma (t)}K_{\sigma (t)})x(t).
 \end{array}
 \end{equation}
 \begin{equation}
 \label{p8_e6}
 \begin{array}{c}
 x(k+1)=(A_{\sigma (k)}+B_{\sigma (k)}K_{\sigma (k)})x(k).
 \end{array}
 \end{equation}

 For the systems (\ref{p8_e5}) and (\ref{p8_e6}), important results were present in some recent papers.

 \begin{lem}\cite{zhao2012stability}(Continuous-time Situation)\label{p8_lem3}
  For the system (\ref{p8_e5}) and the given constants $\lambda_p >0$, $\mu_p>1$, $p\in \rm{{\mathcal S}}$, suppose that there exist matrices $U_p>0$, and  $T_p$ such that $\forall p\in \rm{{\mathcal S}}$,
 	$A_pU_p+B_pT_p+U_pA_p^T+T_p^TB_p^T\le - \lambda_p U_p$
 	and $\forall (p,q)\in {\mathcal S}\times {\mathcal S},p\ne q$,  it holds that
 	$U_q \le \mu_p U_p.$
 	Then there exists a set of stabilizing controllers such that the system (\ref{p8_e5}) is globally uniformly asymptotically stable with the MDADT condition (\ref{p8_e3}). The controller gain can be given by
 	\begin{equation}
 	\label{p8_6}
 	\begin{array}{c}
 	K_p=T_pU_p^{-1}.
 	\end{array}
 	\end{equation}
 	
 \end{lem}

 \begin{lem}\cite{zhao2012stability}(Discrete-time Situation)\label{p8_lem4}
 	For the system (\ref{p8_e6}) and the given constants $1>\lambda_p >0$, $\mu_p\ge1$, $p\in \rm{{\mathcal S}}$, suppose that there exist matrices $U_p>0$, and $T_p$ such that $\forall p\in \rm{{\mathcal S}}$,
 	\[\left[ {\begin{array}{*{20}{c}}
 		{ - {U_p}}&{{A_p}{U_p} + {B_p}{T_p}}\\
 		*&{ - (1 - {\lambda _p}){U_p}}
 		\end{array}} \right] \le 0
 	\]
 	and $\forall (p,q)\in {\mathcal S}\times {\mathcal S},p\ne q$,  it holds that
 	$U_q \le \mu_p U_p.$
 	Then there exists a set of stabilizing controllers such that the system (\ref{p8_e6}) is globally uniformly asymptotically stable with the MDADT condition (\ref{p8_e4}). The controller gain can be given by (\ref{p8_6}).
 	
 \end{lem}

 More detailed contents can be found in recent works \cite{zhang2010asynchronously,hespanha1999stability,zhao2012stability}. By reviewing the existing results about switched systems, we found that the influence of switching sequence on dwell time has not been explicitly considered so far. In the next section of this note, two switching sequence based methods will be proposed.

 \section{MAIN RESULTS}\label{main_results}

 This section includes three subsections. In the first subsection, we propose a sequence-based average subsequence dwell time method. In the second subsection, we  propose a sequence-based average preceding dwell time method. In the third part, we use the two proposed methods to analyze the stability of linear switched systems.

 \subsection{Sequence-based Average Subsequence Dwell Time Method}

 In this subsection, we propose a sequence-based average subsequence dwell time method. According to this method, the setting of the parameter $\mu_p$ or $\mu$ is referred to switching sequences. The relationships among sequences, the stability of switched-systems, average dwell times are reconsidered.

 The definition of the sequence-based average subsequence dwell time is first given as follows.

 \begin{def1}\label{p8_def3}
 	For any switching times $t_2> t_1\ge 0$ with a switching signal $\sigma(t)$, let $N_{\sigma p|q}(t_2,t_1)$ stand for the number of the sequence that the $p^{th}$ subsystem is activated immediately after the $q^{th}$ subsystem over the time interval $[t_1,t_2)$, and let $T_{p,p|q}(t_2,t_1)$ denote the total running time of the $p^{th}$ subsystem activated immediately after the $q^{th}$ subsystems over the time interval $[t_1,t_2)$, $p \in S$. It is said that $\sigma(t)$ has a sequence-based average subsequence dwell time $\tau_{a(p,p|q)}$ if there
 	exist two subsequence numbers $N_{0(p,p|q)}$ ($N_{0(p,p|q)}$ is called the sequenced-based subsequence chatter bound here) and $\tau_{a(p,p|q)}$ such that
 	
 	\begin{equation}\label{p8_e8}
 	N_{\sigma p|q} (t_2,t_1)\le N_{0(p,p|q)} +\frac{T_{p,p|q}(t_2,t_1)}{\tau_{a(p,p|q)} }
 	\end{equation}
 \end{def1}

 \begin{rmk}
 	Definition \ref{p8_def3} constructs a novel set of switching signals referring to the sequence-based average subsequence dwell time (SBASDT). It considers the switching sequence of the switched systems. Obviously, this definition is different from Definition \ref{p8_def2} as the successor of each individual mode has been explicitly considered here.
 \end{rmk}

On the basis of the new Definition \ref{p8_def3}, we present the following two theorems.
\begin{thm}(Continuous-time Situation)\label{p8_thm1}
 	Consider a switched system described by
 	\begin{equation}\label{p8_e9}
 	\dot{x}(t)=f_{\sigma (t)}(x(t)).
 	\end{equation}
Let $\lambda_{q}>0$, $\lambda_{p}>0$, $\mu_{p|q}>1$ be given constants. Suppose there exist $\mathcal C^1$ functions $V_{\sigma(t)}:\mathcal R^n \rightarrow \mathcal R$, and functions $k_{1p}$, $k_{2p}$, $k_{1q}$ and $k_{2q}$ of class  $\mathcal{K}_\infty$, such that $\forall (t_i=p,t_i^-=q)\in {\mathcal S}\times {\mathcal S}$ and $p\neq q$,
 	\begin{equation}
 	\label{p8_e10}	
 	k_{1q}(\|x(t)\|)\le V_q(x(t)) \le k_{2q}(\|x(t)\|),
 	\end{equation}
 	\begin{equation}
 	\label{p8_e11}
 	\dot{V}_q(x(t)) \le -\lambda_{q} V_q(x(t)),
 	\end{equation}
\begin{equation}
 	\label{p8_e12}		
 	k_{1p}(\|x(t)\|)\le V_{p}(x(t)) \le k_{2p}(\|x(t)\|),
 	\end{equation}
 \begin{equation}
 	\label{p8_e13}
 	\dot{V}_{p}(x(t)) \le -\lambda_{p} V_{p}(x(t)),
 	\end{equation}
 and
\begin{equation}
 	\label{p8_e14}
 	V_{p}(x(t_i)) \le \mu_{p|q} V_q(x(t_i)),
 	\end{equation}		
then the system is GUAS for any switching signals with SBASDT
\begin{equation}
 	\label{p8_e15}
 	\tau_{a(p,p|q)} \ge \tau_{a(p,p|q)}^\ast=ln\mu_{p|q}/\lambda_{p}.
 	\end{equation}
\end{thm}
\begin{IEEEproof}
 	Let $t_0=0$ and use $t_1$, $t_2$, $t_3$, $\cdots$ to denote the switching moments. According to (\ref{p8_e13}) and (\ref{p8_e14}), for any time $t>t_0$ and $t\in \left[ t_l,t_{l + 1} \right)$, $l \in Z_+$, we have
 	\begin{equation}
 	\label{p8_e16}
 	\begin{aligned}
 	&\quad  {V_{\sigma (t)}}(x(t))\\
 	&\le \exp \left\{ { - \lambda_{\sigma(t_l)}(t - {t_l})} \right\}{V_{\sigma ({t_l})}}(x({t_l}))\\
 	&\le \exp \left\{ { - \lambda_{\sigma(t_l)}(t - {t_l})} \right\}
 	\mu _{\sigma ({t_l})|\sigma (t_l^ - )} {V_{\sigma ({t_l}^ - )}}(x({t_l}^ - ))\\
 	&\le \exp \left\{ { - {\lambda_{\sigma(t_l)}}(t - {t_l})} \right\}{\mu _{\sigma ({t_l})|\sigma (t_l^ - )}}\\
 	&\quad \times \exp \left\{ { - \lambda_{\sigma(t_{l-1})}({t_l} - {t_{l - 1}})} \right\}{V_{\sigma ({t_{l - 1}})}}(x({t_{l - 1}}))\\
 	&= {\mu _{\sigma ({t_l})|\sigma (t_l^ - )}}\exp \left\{ { - {\lambda_{\sigma ({t_l})}}(t - {t_l})} \right\}\\
 	&\quad \times \exp \left\{ { - {\lambda _{\sigma ({t_{l - 1}})}}({t_l} - {t_{l - 1}})} \right\}{V_{\sigma ({t_{l - 1}})}}(x({t_{l - 1}}))
 	\end{aligned}
 	\end{equation}
Recursively, we have
 	\begin{equation}
 	\label{p8_e17}
 	\begin{small}
 	\begin{array}{l}
 	\quad  {V_{\sigma (t)}}(x(t))\\
 	\le \!\left\{\! {\prod\limits_{k = 1}^l {{\mu _{\sigma ({t_k})|\sigma (t_k^ - )}}} } \!\right\}\!
 \exp \!\left\{\! { \! - \! {\lambda _{\sigma ({t_l})}}(t \! - \! {t_l})} \!  \cdots \!  {\! - \!{\lambda _{\sigma ({t_1})}}({t_2} \! - \! {t_1})} \!\right\}\!\\
 	\quad\times \exp \left\{- \lambda _{\sigma ({0})} ({t_1} - {0})\right\} {V_{\sigma (0)}}(x(0)).
 	\end{array}
 	\end{small}
 	\end{equation}
On the basis of the set ${\mathcal S} = \{ 1,2, \cdots ,s\}$ (where $s$ is the total amount of subsystems), we create a new set ${\mathcal S'}$ which is the Cartesian product of ${\mathcal S}$ and ${\mathcal S}$, i.e., ${\mathcal S'}= {\mathcal S} \times {\mathcal S}=\{(p, q): p \in {\mathcal S}, and \, q \in {\mathcal S} \}$. It is obvious that the set ${\mathcal S'}$ includes all the possible ordered pairs $(\sigma ({t_i}),\sigma ({t_i^-}))$, $i \in Z_+$, and the total number of the elements in ${\mathcal S'}$ is $s' = s(s-1)$.
 	
Consider the system switching up to $l$, and define a new related set $\mathcal S ''=\{(p, q): p \in \sigma ({t_i}), q \in \sigma ({t_i^-}), i=1, 2, 3,\cdots,l\}$. Obviously, $\mathcal S''\subseteq \mathcal S'$.

If $\sigma ({t_n})=p$ and $\sigma (t_n^ - )=q$, we denote $\mu_{\sigma ({t_n})|\sigma (t_n^ - )}$ as $\mu_{p|q}$, $\forall n=1,2,\cdots,l$. The purpose of the formal transformation is to make it more readable under some situations. Furthermore, we list and number all the elements of the set $\mathcal S''$, i.e. we use $ [p|q]_{(k)}$ to denote the pair $(p,q)$ which is the $k^{th}$ element of the set $\mathcal{S}''$. $N_{\sigma [p|q]_{(k)}}(t,0)$ and $T_{p, [p|q]_{(k)}}(t,0)$ denote the activated numbers and total subsequence dwell times of the $k^{th}$ element in the time interval $[0,t)$ respectively. Assuming that the number of elements in ${\mathcal S ''}$ is $s''$, we have $k \in \{1,2,\cdots, s''\}$, $s'' \le s' $, and $s'' \le l $.

Hence,
	\begin{equation}\label{p8_e18}
\begin{array}{l}
\quad V_{\sigma (t)}(x(t))\\
\le \!\left\{\! {\prod\limits_{k = 1}^{s''} {\mu _{[p|q]_{(k)}}^{{N_{\sigma [p|q]_{(k)}}}(t,0)}} } \!\right\}\!
\exp \!{\left\{\! {\sum\limits_{k = 1}^{s''} { - {\lambda _{p,[p|q]_{(k)}}}} {T_{p,[p|q]_{(k)}}}(t,0)} \!\right\}\!}\\
\quad\times \exp \left\{- \lambda _{\sigma ({0})} ({t_1} - {0})\right\}{V_{\sigma (0)}}(x(0)).
\end{array}
\end{equation} 	

On the basis of (\ref{p8_e8}), we can get
 	
\begin{equation}
 	\begin{array}{l}
 	\quad {V_{\sigma (t)}}(x(t))\\
 	\le \left\{ {\prod\limits_{k = 1}^{s''} {\mu _{[p|q]_{(k)}}^{{N_{0[p|q]_{(k)}}} + \frac{{{T_{p,[p|q]_{(k)}}}(t,0)}}{{{\tau _{a(p,[p|q]_{(k)})}}}}}} } \right\}\\
 	\quad \times \exp \left\{ {\sum\limits_{k = 1}^{s''} { - {\lambda _{p,[p|q]_{(k)}}}} {T_{p,[p|q]_{(k)}}}(t,0)} \right\}\\
 	\quad\times \exp \left\{- \lambda _{\sigma ({0})} ({t_1} - {0})\right\}{V_{\sigma (0)}}(x(0))\\
 	\le \exp \left\{ {\sum\limits_{k = 1}^{s''} {{N_{0[p|q]_{(k)}}}\ln {\mu _{[p|q]_{(k)}}}} } \right\}\\
 	\quad \times \exp \left\{ {\sum\limits_{k = 1}^{s''} {\frac{{{T_{p,[p|q]_{(k)}}}(t,0)}}{{{\tau _{a(p,[p|q]_{(k)})}}}}\ln {\mu _{[p|q]_{(k)}}}} } \right\}\\
 	\quad\times \exp \left\{ {\sum\limits_{k = 1}^{s''} { - {\lambda _{p,[p|q]_{(k)}}}} {T_{p,[p|q]_{(k)}}}(t,0)} \right\}\\
 	\quad\times \exp \left\{- \lambda _{\sigma ({0})} ({t_1} - {0})\right\}{V_{\sigma (0)}}(x(0))\\
 	= \exp \left\{ {\sum\limits_{k = 1}^{s''} {{N_{0[p|q]_{(k)}}}\ln {\mu _{[p|q]_{(k)}}}} } \right\}\\
 	\quad\times \exp \left\{ {\sum\limits_{k = 1}^{s''} {(\frac{{\ln {\mu _{[p|q]_{(k)}}}}}{{{\tau _{a(p,[p|q]_{(k)})}}}} \!-\! {\lambda _{p,[p|q]_{(k)}}}){T_{p,[p|q]_{(k)}}}(t,0)} } \right\}\\
 	\quad\times \exp \left\{- \lambda _{\sigma ({0})} ({t_1} - {0})\right\}{V_{\sigma (0)}}(x(0)).
 	\end{array}
 	\end{equation}
 	
 If for all $k \in\{1,2,\cdots,s''\}$
 	\begin{equation}\label{p8_e20}
 	{\tau _{a(p,[p|q]_{(k)})}} \ge \ln {\mu _{[p|q]_{(k)}}}/{\lambda _{p,[p|q]_{(k)}}},
 	\end{equation}
then, we have
\begin{equation}
 	\begin{small}
 	\begin{array}{l}
 	\quad {V_{\sigma (t)}}(x(t))\\
 	\le \exp \left\{ {\sum\limits_{k = 1}^{s''} {{N_{0[p|q]_{(k)}}}\ln {\mu _{[p|q]_{(k)}}}} } \right\}\\
 	\quad \times \exp \!\left\{\! {\mathop {\max }\limits \!\left\{\! {\mathop {\max }\limits_k \!\left\{\! {\frac{{\ln {\mu _{[p|q]_{(k)}}}}}{{{\tau _{a(p,[p|q]_{(k)})}}}} \!-\! {\lambda _{p,[p|q]_{(k)}}}} \!\right\}\!, \!-\! {\lambda _{\sigma({0})}}} \!\right\}\!t} \!\right\}\!\\
 	\quad \times {V_{\sigma (0)}}(x(0)).
 	\end{array}
 	\end{small}
 	\end{equation}
 	
 	Therefore, we conclude that ${V_{\sigma (t)}}(x(t))$ convergences to zero as $t \to  + \infty $ if the SBASDT satisfies the condition (\ref{p8_e15}). Then, the asymptotic stability can be deduced according to Definition \ref{p8_def1}.
 \end{IEEEproof}

 \begin{thm}(Discrete-time Situation)\label{p8_thm2}
 Consider a discrete switched system described by
 	\begin{equation}\label{p8_e22}
 	x(k+1)=f_{\sigma (k)}(x(k)),
 	\end{equation}
and let $1>\lambda_{q}>0$, $1>\lambda_{p}>0$, $\mu_{p|q}>1$ be given constants. Assume there exist $\mathcal C^1$ functions $V_{\sigma(k)}:\mathcal R^n \rightarrow \mathcal R$ and functions $k_{1p}$, $k_{2p}$, $k_{1q}$ and $k_{2q}$ of class $\mathcal{K}_\infty$, such that $\forall (k_i=p,k_i-1=q)\in {\mathcal S}\times {\mathcal S}$, $p\neq q$,
\begin{equation}
 	\label{p8_e23}	
 	k_{1q}(\|x(k)\|)\le V_q(x(k)) \le k_{2q}(\|x(k)\|),
 	\end{equation}
\begin{equation}
 	\label{p8_e24}
 	V_q(x(k+1))-V_q(x(k)) \le -\lambda_{q} V_q(x(k)),
 	\end{equation}
\begin{equation}
 	\label{p8_e25}		
 	k_{1p}(\|x(k)\|)\le V_{p}(x(k)) \le k_{2p}(\|x(k)\|),
 	\end{equation}
\begin{equation}
 	\label{p8_e26}
 	V_{p}(x(k+1))-V_{p}(x(k)) \le -\lambda_{p} V_{p}(x(k)),
 	\end{equation}
and
\begin{equation}
 	\label{p8_e27}
 	V_{p}(x(k_i)) \le \mu_{p|q} V_q(x(k_i)),
 	\end{equation}	
 then the system is GUAS for any switching signals with SBASDT
 \begin{equation}
 	\label{p8_e28}
 	\tau_{a(p,p|q)} \ge
 	\tau_{a(p,p|q)}^\ast=-ln\mu_{p|q}/ln(1-\lambda_{p}).
 	\end{equation}
 \end{thm}
\begin{IEEEproof}
This theorem can be proved by using a method similar to the proof of the continue-time situation. Due to the space limitation, detailed proof is omitted here.
 \end{IEEEproof}

 \begin{rmk}
Actually, the conditions (\ref{p8_e10})-(\ref{p8_e13}) (or (\ref{p8_e23})-(\ref{p8_e26})) are equivalent to the corresponding conditions of the MDADT method \cite{zhao2012stability}. This remark is also valid for the sequence-based average preceding dwell time method.
\end{rmk}
\begin{rmk}
 Compared to the existing results, new methods allow a subsystem to have different switching restrictions $\mu_{p|q}$ for different sequences.
\end{rmk}
\begin{rmk}
  MDADT is a popular method for the analysis of the switched systems. For the $p^{th}$ subsystem, if we set the values of $\mu_{p|q}$ as a fixed value $\mu_{p}$, without considering the preceding different $q^{th}$ subsystems, the SBASDT retrogresses to MDADT. Therefore, the method in this note has less conservativeness and restriction than MDADT.
 \end{rmk}
\subsection{Sequence-based Average Preceding Dwell Time Method}
In this subsection, we propose another sequence-based method, the sequence-based average preceding dwell time method.

The definition of the sequence-based average preceding dwell time is given as follows.
 \begin{def1}\label{p8_def4}
 	For any switching times $t_2> t_1\ge 0$ with a switching signal $\sigma(t)$, let $N_{\sigma p|q}(t_2,t_1)$ stand for the number of the sequence that the $p^{th}$ subsystem is activated immediately after the $q^{th}$ subsystem over the time interval $[t_1,t_2)$, and let $T_{q,p|q}(t_2,t_1)$ denote the total running time of the $q^{th}$ subsystem when the $p^{th}$ subsystem is activated immediately after the $q^{th}$ subsystems over the time interval $[t_1,t_2)$, $p \in S$.
 It is said that $\sigma(t)$ has a sequence-based average preceding dwell time $\tau_{a(q,p|q)}$ if there exist two preceding numbers $N_{0(q,p|q)}$ ($N_{0(q,p|q)}$ is called the sequenced-based preceding chatter bound here) and $\tau_{a(q,p|q)}$ such that
 	
 	\begin{equation}\label{p8_e29}
 	N_{\sigma p|q} (t_2,t_1)\le N_{0(q,p|q)} +\frac{T_{q,p|q}(t_2,t_1)}{\tau_{a(q,p|q)} }
 	\end{equation}
 \end{def1}

 \begin{rmk} 	
 	Definition \ref{p8_def4} also constructs a novel set of switching signals referring to the sequence-based average preceding dwell time method (SBAPDT). Both Definition \ref{p8_def3} and Definition \ref{p8_def4} are based on switching sequence.
 \end{rmk}

 On the basis of Definition \ref{p8_def4}, we can get the following two theorems.

 \begin{thm}(Continuous-time Situation)\label{p8_thm3}
 Consider a switched system depicted by (\ref{p8_e9}),
 and let $\lambda_{q}>0$, $\lambda_{p}>0$, $\mu_{p|q}>1$ be given constants. Suppose there exist $\mathcal C^1$ functions $V_{\sigma(t)}:\mathcal R^n \rightarrow \mathcal R$, and functions $k_{1p}$, $k_{2p}$, $k_{1q}$ and $k_{2q}$ of class $\mathcal{K}_\infty$, such that $\forall (t_i=p,t_i^-=q)\in {\mathcal S}\times {\mathcal S}$, $p\neq q$,
  	\begin{equation}
 	\label{p8_e30}	
 	k_{1q}(\|x(t)\|)\le V_q(x(t)) \le k_{2q}(\|x(t)\|),
 	\end{equation}
 \begin{equation}
 	\label{p8_e31}
 	\dot{V}_q(x(t)) \le -\lambda_{q} V_q(x(t)),
 	\end{equation}
 \begin{equation}
 	\label{p8_e32}		
 	k_{1p}(\|x(t)\|)\le V_{p}(x(t)) \le k_{2p}(\|x(t)\|),
 	\end{equation}
 \begin{equation}
 	\label{p8_e33}
 	\dot{V}_{p}(x(t)) \le -\lambda_{p} V_{p}(x(t)),
 	\end{equation}
  	and
  	\begin{equation}
 	\label{p8_e34}
 	V_{p}(x(t_i)) \le \mu_{p|q} V_q(x(t_i)),
 	\end{equation}		
 then the system is GUAS for any switching signals with SBAPDT
  	\begin{equation}
 	\label{p8_e35}
 	\tau_{a(q,p|q)} \ge ln\mu_{p|q}/\lambda_{q}.
 	\end{equation}
  \end{thm}
\begin{IEEEproof}
 Let $t_0=0$ and use $t_1$, $t_2$, $t_3$, $\cdots$ to denote the subsystems switching times. According to (\ref{p8_e33}) and (\ref{p8_e34}), for any time $t>t_0$ and $t\in \left[ t_l,t_{l + 1} \right)$, $l \in Z_+$, we have systems (\ref{p8_e16})-(\ref{p8_e17}).
 	
 We also use the definitions of ${\mathcal S}$, ${\mathcal S'}$, ${\mathcal S''}$, and $[p|q]_{(k)}$ from the proof of Theorem \ref{p8_thm1}. The symbol $T_{q, [p|q]_{(k)}}(t,0)$ denotes the total dwell times of the $q^{th}$subsystem for the $k^{th}$ sequence union in the time interval $[0,t)$.
 	
 	Hence,	
 	\begin{equation}\label{p8_e36}
 \begin{array}{l}
 \quad {V_{\sigma (t)}}(x(t))\\
 \le \left\{ {\prod\limits_{k = 1}^{s''} {\mu _{[p|q]_{(k)}}^{{N_{\sigma [p|q]_{(k)}}}(t,0)}} } \right\}\\
 \quad\times \exp \left\{ { {\sum\limits_{k = 1}^{s''} { - {\lambda _{q,[p|q]_{(k)}}}} {T_{q,[p|q]_{(k)}}}(t,0)} } \right\}\\
 \quad\times \exp \left\{- \lambda _{\sigma ({t_l})} ({t} - {t_l})\right\}{V_{\sigma (0)}}(x(0)).
 \end{array}
 \end{equation}	
 	According to (\ref{p8_e29}), one can get
 	
 	\begin{equation}
 	\begin{array}{l}
 	\quad {V_{\sigma (t)}}(x(t))\\
 	\le \left\{ {\prod\limits_{k = 1}^{s''} {\mu _{[p|q]_{(k)}}^{{N_{0(q,[p|q]_{(k)})}} + \frac{{{T_{q,[p|q]_{(k)}}}(t,0)}}{{{\tau _{a(q,[p|q]_{(k)})}}}}}} } \right\}\\
 	\quad \times \exp \left\{ {\sum\limits_{k = 1}^{s''} { - {\lambda _{q,[p|q]_{(k)}}}} {T_{q,[p|q]_{(k)}}}(t,0)} \right\}\\
 	\quad\times \exp \left\{- \lambda _{\sigma ({t_l})} ({t} - {t_l})\right\} {V_{\sigma (0)}}(x(0))\\
 	\le \exp \left\{ {\sum\limits_{k = 1}^{s''} {{N_{0(q,[p|q]_{(k)})}}\ln {\mu _{[p|q]_{(k)}}}} } \right\}\\
 	\quad \times \exp \left\{ {\sum\limits_{k = 1}^{s''} {\frac{{{T_{q,[p|q]_{(k)}}}(t,0)}}{{{\tau _{a(q,[p|q]_{(k)})}}}}\ln {\mu _{[p|q]_{(k)}}}} } \right\}\\
 	\quad\times \exp \left\{ {\sum\limits_{k = 1}^{s''} { - {\lambda _{q,[p|q]_{(k)}}}} {T_{q,[p|q]_{(k)}}}(t,0)} \right\}\\
 	\quad\times \exp \left\{- \lambda _{\sigma ({t_l})} ({t} - {t_l})\right\}{V_{\sigma (0)}}(x(0))\\
 	= \exp \left\{ {\sum\limits_{k = 1}^{s''} {{N_{0(q,[p|q]_{(k)})}}\ln {\mu _{[p|q]_{(k)}}}} } \right\}\\
 	\quad\times \exp \left\{ {\sum\limits_{k = 1}^{s''} {(\frac{{\ln {\mu _{[p|q]_{(k)}}}}}{{{\tau _{a(q,[p|q]_{(k)})}}}} - {\lambda _{q,[p|q]_{(k)}}}){T_{q,[p|q]_{(k)}}}(t,0)} } \! \right\}\!\\
 	\quad\times \exp \left\{- \lambda _{\sigma ({t_l})} ({t} - {t_l})\right\}{V_{\sigma (0)}}(x(0)).
 	\end{array}
 	\end{equation}
 	
 	If there exist constants
 	\[
 	{\tau _{a(q,[p|q]_{(k)})}} \ge \ln {\mu _{[p|q]_{(k)}}}/{\lambda _{q,[p|q]_{(k)}}},
 	\]
we have
	\begin{equation}
 	\begin{small}
 	\begin{array}{l}
 	\quad {V_{\sigma (t)}}(x(t))\\
 	\le \exp \left\{ {\sum\limits_{k = 1}^{s''} {{N_{0(q,[p|q]_{(k)})}}\ln {\mu _{[p|q]_{(k)}}}} } \right\}\\
 	\quad\times \exp \!\left\{\! {\mathop {\max }\limits \!\left\{\! {\mathop {\max }\limits_k \!\left\{\! {\frac{{\ln {\mu _{[p|q]_{(k)}}}}}{{{\tau _{a(q,[p|q]_{(k)})}}}} \!-\! {\lambda _{q,[p|q]_{(k)}}}} \right\}, - {\lambda _{\sigma (t_l)}}} \right\}t} \right\}\\
 	\quad \times {V_{\sigma (0)}}(x(0))
 	\end{array}
 	\end{small}
 	\end{equation}
Therefore, we conclude that ${V_{\sigma (t)}}(x(t))$ convergences to zero as $t \to  + \infty $ if the SBAPDT satisfies the condition (\ref{p8_e35}). Then, the asymptotic stability can be deduced according to Definition \ref{p8_def1}.
 \end{IEEEproof}
\begin{thm}\label{p8_thm4}(Discrete-time Situation)
Consider a discrete switched system described by (\ref{p8_e22}),
 	and let $1>\lambda_{q}>0$, $1>\lambda_{p}>0$, $\mu_{p|q}>1$ be given constants. Assume there exist $\mathcal C^1$ functions $V_{\sigma(k)}:\mathcal R^n \rightarrow \mathcal R$, and functions $k_{1p}$, $k_{2p}$, $k_{1q}$ and $k_{2q}$ of  class $\mathcal K_\infty$, such that $\forall (k_i=p,k_i-1=q)\in {\mathcal S}\times {\mathcal S}$, $p\neq q$,
 \begin{equation}
 	\label{p8_e39}	
 	k_{1q}(\|x(k)\|)\le V_q(x(k)) \le k_{2q}(\|x(k)\|),
 	\end{equation}
\begin{equation}
 	\label{p8_e40}
 	V_q(x(k+1))-V_q(x(k)) \le -\lambda_{q} V_q(x(k)),
 	\end{equation}
\begin{equation}
 	\label{p8_e41}		
 	k_{1p}(\|x(k)\|)\le V_{p}(x(k)) \le k_{2p}(\|x(k)\|),
 	\end{equation}
\begin{equation}
 	\label{p8_e42}
 	V_{p}(x(k+1))-V_{p}(x(k)) \le -\lambda_{p} V_{p}(x(k)),
 	\end{equation}
and
\begin{equation}
 	\label{p8_e43}
 	V_{p}(x(k_i)) \le \mu_{p|q} V_q(x(k_i)),
 	\end{equation}	
then the system is GUAS for any switching signals with SBAPDT
\begin{equation}
 	\label{p8_e44}
 	\tau_{a(q,p|q)} \ge -ln\mu_{p|q}/ln(1-\lambda_{q}).
 	\end{equation}
\end{thm}
\begin{IEEEproof}
This theorem can be proved by using a method similar to the proof of the continue-time situation. To save space, the proof is omitted here.
 \end{IEEEproof}

 \subsection{Linear Switched Systems with two Sequence-based Average Dwell Time Methods}

 In this subsection, we study the stability of switched systems using SBASDT and SBAPDT.

 First, we consider the general linear system (\ref{p8_e1}) and (\ref{p8_e2}). When $u(t)\equiv0$, we can obtain Corollary \ref{p8_cor1} and \ref{p8_cor2} below.
\begin{cor1}\label{p8_cor1}(Continuous-time Situation)
Consider a switched system depicted by (\ref{p8_e1}) with $u(t)\equiv0$. Let $\lambda_{q}>0$, $\lambda_{p}>0$, $\mu_{p|q}>1$ be given constants. Suppose that there exist matrices $P_{q}>0$, $P_{p}>0$, $p,q\in \rm{{\mathcal S}}$, such that $\forall p\in \rm{{\mathcal S}}$, $\forall (t_i=p,t_i^-=q)\in {\mathcal S}\times {\mathcal S}$, $p\neq q$,
\begin{equation}
 	A_{q}^TP_{q}+P_{q}A_{q} \le -\lambda_{q} P_{q},
 	\end{equation}
\begin{equation}
 	A_{p}^TP_{p}+P_{p}A_{p} \le -\lambda_{p} P_{p},
 	\end{equation}
and
\begin{equation}
 	P_{p} \le \mu_{p|q} P_{q},
 	\end{equation}		
then the system is GUAS for any switching signals with SBASDT
\begin{equation}
 	\label{p8_e48}
 	\tau_{a(p,p|q)} \ge \tau_{a(p,p|q)}^\ast=ln\mu_{p|q}/\lambda_{p}.
 	\end{equation}	
 	or with SBAPDT
\begin{equation}
 	 \label{p8_e49}
 	\tau_{a(q,p|q)} \ge \tau_{a(q,p|q)}^\ast=ln\mu_{p|q}/\lambda_{q}.
 	\end{equation}	
\end{cor1}
\begin{IEEEproof}
This corollary can be proved according to Theorems \ref{p8_thm1} and \ref{p8_thm3} and the proof of Lemma \ref{p8_lem1}. Detailed proof is omitted here.
\end{IEEEproof}
\begin{cor1}\label{p8_cor2}(Discrete-time Situation)
Consider a discrete switched system described by (\ref{p8_e2})
with $u(t)\equiv0$.
Let $1>\lambda_{q}>0$, $1>\lambda_{p}>0$, $\mu_{p|q}>1$ be given constants. Suppose that there exist matrices $P_{q}>0$, $P_{p}>0$, $p,q\in \rm{{\mathcal S}}$, such that $\forall (k_i=p,k_i-1=q)\in {\mathcal S}\times {\mathcal S}$, $p\neq q$,
\begin{equation}
 	A_{q}^TP_{q}A_{q}-P_{q}\le - \lambda_{q} P_{q}
 	\end{equation}
\begin{equation}		
 	A_{p}^TP_{p}A_{p}-P_{p}\le - \lambda_{p} P_{p}
 	\end{equation}
and
\begin{equation}
 	P_{p} \le \mu_{p|q} P_{q},
 	\end{equation}	
then the system is GUAS for any switching signals with SBASDT
\begin{equation}
 	 	\label{p8_e53}
 	\tau_{a(p,p|q)} \ge -ln\mu_{p|q}/ln(1-\lambda_{p}).
 	\end{equation}
 	or with SBAPDT
 	\begin{equation}
 	 	\label{p8_e54}
 	\tau_{a(q,p|q)} \ge -ln\mu_{p|q}/ln(1-\lambda_{q}).
 	\end{equation}
 \end{cor1}
\begin{IEEEproof}
	This corollary can be proved by using a method similar to the proof of the continue-time situation. Detailed proof is omitted here.
\end{IEEEproof}
\begin{table*}[!t]\centering
 	\caption{\label{table1}compution for the MDADT and SBASDT switching}
 	\begin{tabular}{|l||l|l|}
 		\hline
 		Schemes & MDADT switching & SBASDT  switching \\
 		\hline
 		Criteria & Theorem 1 in \cite{zhang2010asynchronously} & Corollary \ref{p8_cor5} in this note \\\hline
 		Parameters & $\lambda_1$=3; $\lambda_2=1.5$; $\lambda_3=2.5$; & $\lambda_1$=3; $\lambda_2=1.5$;  $\lambda_3=2.5$;\\
 		\quad & $\mu_1=18$; $\mu_2=2.3$; $\mu_3=41.$ &$\mu_{1|2}=18$; $\mu_{2|1}=2.3$; $\mu_{3|1}=41$;\\
 		\quad & \quad &$\mu_{1|3}=13$; $\mu_{2|3}\rightarrow1$; $\mu_{3|2}=17$.\\	\hline
 		\quad & $K1 =[403.6393, -107.2597]$; &K1 =[371.7662, -100.0154]； \\
 		Controller & $K2 =[2.8646,   -0.6579]; $&K2 =[2.8330,   -0.4917]； \\
 		\quad &$K3 =[-9.4055,   -2.6831]$.  & 	K3 =[-7.9922,   -2.0744].\\ \hline
 		Average & $\tau_{a1}^\ast$=0.96; & $\tau_{a(1,1|2)}^\ast=0.96$; $\tau_{a(1,1|3)}^\ast=0.86$; \\
 		Dwell Time& $\tau_{a2}^\ast$=0.56; &$\tau_{a(2,2|1)}^\ast=0.56$; $\tau_{a(2,2|3)}^\ast\rightarrow0$;\\
 		Thresholds & $\tau_{a3}^\ast$=1.5. &  $\tau_{a(3,3|2)}^\ast=1.1$; $\tau_{a(3,3|1)}^\ast=1.5$.\\
 		\hline
 	\end{tabular}
 \end{table*}
\begin{table*}[!t]\centering
 	\caption{\label{table2}dwell times of the SBASDT and SBAPDT switching}
 	\begin{tabular}{|l||l|l|}
 		\hline
 		Schemes & SBASDT switching & SBAPDT  switching \\
 		\hline
 		Average &  $\tau_{a(1,1|2)}^\ast=0.96$; $\tau_{a(1,1|3)}^\ast=0.86$; & $\tau_{a(2,1|2)}^\ast=1.9$; $\tau_{a(3,1|3)}^\ast=1.0$;\\
 		Dwell Time& $\tau_{a(2,2|1)}^\ast=0.56$; $\tau_{a(2,2|3)}^\ast\rightarrow0$;&$\tau_{a(1,2|1)}^\ast=0.28$; $\tau_{a(3,2|3)}^\ast\rightarrow0$;\\
 		Thresholds&  $\tau_{a(3,3|2)}^\ast=1.1$; $\tau_{a(3,3|1)}^\ast=1.5$.&  $\tau_{a(2,3|2)}^\ast=1.9$; $\tau_{a(1,3|1)}^\ast=1.2$.\\
 		\hline
 	\end{tabular}
 \end{table*}

For the case $u(t)\not\equiv 0$,  we have Corollaries \ref{p8_cor3} and \ref{p8_cor4}. Their proofs are very similar to those of Corollaries \ref{p8_cor1} and \ref{p8_cor2}, respectively, and hence are omitted.
\begin{cor1}(Continuous-time Situation)\label{p8_cor3}
Consider a switched system depicted by (\ref{p8_e1})
with $u(t)=K_{\sigma(t)}x(t)$.
Let $\lambda_{q}>0$, $\lambda_{p}>0$, $\mu_{p|q}>1$ be given constants.  Suppose that there exist matrices $P_{q}>0$, $P_{p}>0$, $p,q\in \rm{{\mathcal S}}$, such that $\forall (t_i=p,t_i^-=q)\in {\mathcal S}\times {\mathcal S}$, $p\neq q$,
\begin{equation}
 	\begin{array}{l}
 	(A_{q}+B_{q}K_{q})^TP_{q}
 	+P_{q}(A_{q}+B_{q}K_{q})
 	\le -\lambda_{q} P_{q},
 	\end{array}
 	\end{equation}
\begin{equation}
 	\begin{array}{l}
 	(A_{p}+B_{p}K_{p})^TP_{p}
 	+P_{p}(A_{p}+B_{p}K_{p})
 	\le -\lambda_{p} P_{p},
 	\end{array}
 	\end{equation}
and
\begin{equation}
 	P_{p} \le \mu_{p|q} P_{q},
 	\end{equation}		
 	then the system is GUAS for any switching signals with the SBASDT switching condition (\ref{p8_e48})	
 	or with the SBAPDT switching condition (\ref{p8_e49}).
 \end{cor1}
\begin{cor1}\label{p8_cor4}(Discrete-time Situation)
 	For a discrete switched system
  (\ref{p8_e2})
 	and  $u(k)=K_{\sigma(k)}x(t)$, let $1>\lambda_{q}>0$, $1>\lambda_{p}>0$, $\mu_{p|q}>1$ be given constants.  Suppose that there exist matrices $P_{q}>0$, $P_{p}>0$, $p,q\in \rm{{\mathcal S}}$, such that $\forall (k_i=p,k_i-1=q)\in {\mathcal S}\times {\mathcal S}$, $p\neq q$
\begin{equation}
 	\begin{array}{l}
 	(A_{q}\!+\!B_{q}K_{q})^TP_{q}(A_{q}\!+\!B_{q}K_{q})
 	\!-\!P_{q}
 	\le - \lambda_{q} P_{q}
 	\end{array}
 	\end{equation}
\begin{equation}
 	\begin{array}{l}
 	(A_{p}+B_{p}K_{p})^TP_{p}(A_{p}+B_{p}K_{p})
 	-P_{p}
 	\le - \lambda_{p} P_{p}
 	\end{array}
 	\end{equation}	
 	and
 	\begin{equation}
 	P_{p} \le \mu_{p|q} P_{q},
 	\end{equation}	
then the system is GUAS for any switching signals with the SBASDT switching condition (\ref{p8_e53})
 	or with the SBAPDT switching condition. (\ref{p8_e54}).
 \end{cor1}

If we want to get a solution on controller gain $K_{\sigma(t)}$ and $K_{\sigma(k)}$, the following two corollaries can provide solutions.
 \begin{cor1}\label{p8_cor5}(Continuous-time Situation)
 Consider a switched system depicted by (\ref{p8_e1}), and let $\lambda_{q}>0$, $\lambda_{p}>0$, $\mu_{p|q}>1$ be given constants. Suppose that there exist matrices $U_p>0$, and $T_p$, $p\in \rm{{\mathcal S}}$, such that $\forall (t_i=p,t_i^-=q)\in {\mathcal S}\times {\mathcal S}$, $p\neq q$,
\begin{equation}
 	\begin{array}{l}
 	A_pU_p+B_pT_p+U_pA_p^T+T_p^TB_p^T\le - \lambda_p U_p
 	\end{array}
 	\end{equation}
\begin{equation}
 	\begin{array}{l}
 	A_qU_q+B_qT_q+U_qA_q^T+T_q^TB_q^T\le - \lambda_q U_q
 	\end{array}
 	\end{equation}
and
\begin{equation}
 	U_q \le \mu_{p|q} U_p.
 	\end{equation}		
Then there exists a set of stabilizing controllers such that the system is globally uniformly asymptotically stable with the  with SBASDT (\ref{p8_e48}) or with SBAPDT condition (\ref{p8_e49}).
 	
The controller gain can be calculated by
\begin{equation}
 	\label{p8_e68}
 	\begin{array}{c}
 	K_p=T_pU_p^{-1}.
 	\end{array}
 	\end{equation}
\begin{equation}
 	\label{p8_e69}
 	\begin{array}{c}
 	K_q=T_qU_q^{-1}.
 	\end{array}
 	\end{equation}
\end{cor1}
\begin{IEEEproof}
This corollary can be established according to Corollary \ref{p8_cor3} and the proof of lemma \ref{p8_lem3}. Detailed proof is omitted here.
\end{IEEEproof}
\begin{cor1}\label{p8_cor6}(Discrete-time Situation)
Consider a discrete switched system described by (\ref{p8_e2}), and let $1>\lambda_{q}>0$, $1>\lambda_{p}>0$, $\mu_{p|q}>1$ be given constants. Suppose that there exist matrices $U_{q}>0$, $U_{p}>0$, $T_p$, $T_q$, $p,q\in \rm{{\mathcal S}}$, such that $\forall (k_i=p,k_i-1=q)\in {\mathcal S}\times {\mathcal S}$, $p\neq q$,  	
\begin{equation}
 	\left[ {\begin{array}{*{20}{c}}
 		{ - {U_p}}&{{A_p}{U_p} + {B_p}{T_p}}\\
 		*&{ - (1 - {\lambda _p}){U_p}}
 		\end{array}} \right] \le 0
 	\end{equation}
\begin{equation}
 	\left[ {\begin{array}{*{20}{c}}
 		{ - {U_q}}&{{A_q}{U_q} + {B_q}{T_q}}\\
 		*&{ - (1 - {\lambda _q}){U_q}}
 		\end{array}} \right] \le 0
 	\end{equation}
and
\begin{equation}
 	U_{q} \le \mu_{p|q} U_{p},
 	\end{equation}	
Then there exists a set of stabilizing controllers such that the system is globally uniformly asymptotically stable with the SBASDT condition (\ref{p8_e53}) or with the SBAPDT condition (\ref{p8_e54}).
 	
The controller gain can be designed through (\ref{p8_e68})-(\ref{p8_e69}).
\end{cor1}
\begin{IEEEproof}
This corollary can be derived according to Corollary \ref{p8_cor4} and the proof of lemma \ref{p8_lem4}. Detailed proof is omitted here.
\end{IEEEproof}

\section{NUMERICAL EXAMPLE}
In this section, we use a numerical example to show the effectiveness of the proposed methods and the derived results. Because of the space limitation and the similarity between continuous-time and discrete-time systems, we only verify Corollary \ref{p8_cor5}. Other obtained results can be verified in the same way.

Consider switched linear systems including three subsystems described as

$A_1=\left[ {\begin{array}{*{20}{c}}
 	100.3&-20.1\\
 	-10.1& -10.2
 	\end{array}} \right]$,
 $B_1=\left[ {\begin{array}{*{20}{c}}-0.5&-0.8
 	\end{array}} \right]$,

 $A_2=\left[ {\begin{array}{*{20}{c}}
 	10.8&-10.2\\
 	2& 10.5
 	\end{array}} \right]$,
 $B_2=\left[ {\begin{array}{*{20}{c}}-10.1&10
 	\end{array}} \right]$,

 $A_3=\left[ {\begin{array}{*{20}{c}}
 	0.2&-3.59\\
 	12& 10.4
 	\end{array}} \right]$,
 $B_3=\left[ {\begin{array}{*{20}{c}}5.1&-10.1
 	\end{array}} \right]$,

In this example, we want to get a set of proper sequence-based stabilizing controller gains and search the admissible minimal SBADT to make the closed-loop system stable.

In order to show the advantage of the sequence-based average dwell time, we use Table \ref{table1} to show the numerical results for the mode-dependent average dwell time switching \cite{zhao2012stability} and the sequence-based average subsequence dwell time switching.

It can be seen that the dwell time thresholds are reduced notably according to Table \ref{table1}. For the first dwell time set, $0.96$ and 0.86 repalce 0.96. Two values with one being 0.56 and the other approaching 0 replace 0.56 for the second set. The values 1.1 and 1.5 replace 1.5 for the third set.The control gains are also shown in Table \ref{table1}. The dwell times of two sequence-based methods are shown in Table \ref{table2}. We take the first sequence, $1|2$, in Table \ref{table2} as an example. According to the conditions (\ref{p8_e48}) and (\ref{p8_e49}), we have $\tau_{a(1,1|2)}^\ast=0.96$ and $\tau_{a(2,1|2)}^\ast=1.9$. The value $\tau_{a(1,1|2)}^\ast$ is smaller than $\tau_{a(2,1|2)}$ for $\lambda_1>\lambda_2$. For the same reason, $\tau_{a(2,2|1)}^\ast>\tau_{a(1,2|1)}^\ast$.

 \section{CONCLUSION }
In this note, considering the switching sequence of subsystems, we have proposed
two new analysis methods, for studying the stability of the switched systems. The proposed sequence based average dwell time switching is less conservative than the mode-dependent average dwell time switching, as well as the traditional average dwell time switching.
Both linear and non-linear systems in continuous-time and discrete-time situations were analyzed, and the associated state feedback controllers were designed. Finally, we presented an illustrative example to show the advantages and effectiveness of the proposed new methods.


\end{document}